\documentclass{article}
\usepackage{arxiv}
\usepackage[utf8]{inputenc} 
\usepackage[T1]{fontenc}    
\usepackage{hyperref}
\usepackage{ upgreek }
\usepackage{url}            
\usepackage{booktabs}       
\usepackage{amsfonts}       
\usepackage{nicefrac}       
\usepackage{microtype}      
\usepackage{lipsum}
\usepackage{float}
\usepackage{graphicx}
\usepackage{multirow}
\usepackage{array}
\usepackage{comment}
\usepackage{longtable}
\usepackage{ragged2e}
\usepackage{subfig}
\usepackage{amsmath}
\usepackage{longtable}
\usepackage{ amssymb }
\usepackage{caption}

\title{Online learning for X-ray, CT or MRI}

\author{
  Mosabbir Bhuiyan   \\
  Research and Development Department, Pioneer Alpha\\
  Dhaka, Bangladesh\\
  \texttt{mosabbirbhuiyanonady@gmail.com} \\
   \And
    MD Abdullah Al Nasim   \\
 Research and Development Department, Pioneer Alpha\\
  Dhaka, Bangladesh\\
  \texttt{nasim.abdullah@ieee.org} \\
   \And
      Sarwar Saif \\
   Research and Development Department, Pioneer Alpha\\
  Dhaka, Bangladesh\\
  \texttt{saifmu6@gmail.com} \\
   \And
  Dr. Kishor Datta Gupta \\
 Clark Atlanta University\\
  Georgia, USA\\
  \texttt{kgupta@cau.edu} \\
   \And
   Md Jahangir Alam \\
  Department of Computer Science\\
  University of Alabama at Birmingham, Alabama, USA\\
  \texttt{malam@uab.edu}\\
  \And
 Sajedul Talukder \\
  Department of Computer Science\\
  University of Alabama at Birmingham, Alabama, USA\\
  \texttt{stalukder@uab.edu}}

\begin{document}
\maketitle

\begin{abstract}
Medical imaging plays an important role in the medical sector in identifying diseases. X-ray, computed tomography (CT) scans, and magnetic resonance imaging (MRI) are a few examples of medical imaging. Most of the time, these imaging techniques are utilized to examine and diagnose diseases. Medical professionals identify the problem after analyzing the images. However, manual identification can be challenging because the human eye is not always able to recognize complex patterns in an image. Because of this, it is difficult for any professional to recognize a disease with rapidity and accuracy. In recent years, medical professionals have started adopting Computer-Aided Diagnosis (CAD) systems to evaluate medical images. This system can analyze the image and detect the disease very precisely and quickly. However, this system has certain drawbacks in that it needs to be processed before analysis. Medical research is already entered a new era of research which is called Artificial Intelligence (AI). AI can automatically find complex patterns from an image and identify diseases. Methods for medical imaging that uses AI techniques will be covered in this chapter.

\end{abstract}




\maketitle

\section{Introduction}\label{sec1}

\begin{figure}
\centering
\includegraphics[height=6.2cm]{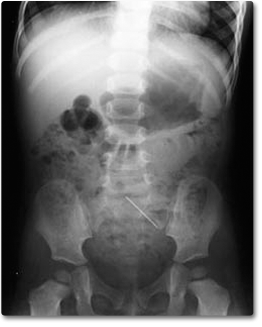}
\caption{An X-ray photo of a one-year-old girl who swallowed a sewing pin~\cite{100}.}
\label{fig:fig_xrays}
\end{figure}

Many people pass away every year because of substandard care and facilities. The lives of many people could be saved if a disease is discovered early. Therefore, a branch called radiology is introduced in the field of medicine that uses imaging techniques to cure and diagnose disorders~\cite{novelline2004squire,chen2011basic,herring2019learning}. Diagnostic radiology and interventional radiology are two subcategories of radiology. Diagnostic radiology allows radiologists viewing of internal body structures. In this area, radiologists identify the underlying source of symptoms, track how effectively the body responds to treatment, and do disease screenings. The most prevalent diagnostic radiology procedures include computed tomography (CT), magnetic resonance imaging (MRI) and magnetic resonance angiography (MRA), mammography, x-ray, positron emission tomography (PET) scan, ultrasound, etc. Interventional radiologists employ imaging techniques like CT and MRI to help direct treatments. Doctors can insert catheters, wires, or other small instruments into your body with the use of imaging. Angiography, cancer treatment, needle organ biopsies, uterine artery embolization, etc are some examples of interventional radiology techniques~\cite{105}.
X-rays are a form of electromagnetic radiation where the wavelength is ranging from 10 picometers to 10 nanometers, corresponding to frequencies in the range of 30 petahertz to 30 exahertz. Wavelengths of X-rays are generally longer compared to those of gamma rays but shorter compared to those of UV rays. In 1895, German scientist Wilhelm Conrad Roentgen first discovered and documented X-rays. An X-ray is a rapid, painless diagnostic that generates pictures of the internal organs and structures in your body, especially your bones. Depending on the density of the parts, the body absorbs X-ray rays when they pass through it. Dense substances like bone and metal appear white when an X-ray is observed. Muscle and fat appear as shades of gray, whereas the air in the lungs is black. In the medical field, X-rays can be used to find fractures and infections in bones or teeth, bone cancer, lung infections, breast cancer, blocked blood vessels, etc. Figure~\ref{fig:fig_xrays} shows an x-ray image.

\begin{figure}
\centering
\includegraphics[height=6.2cm]{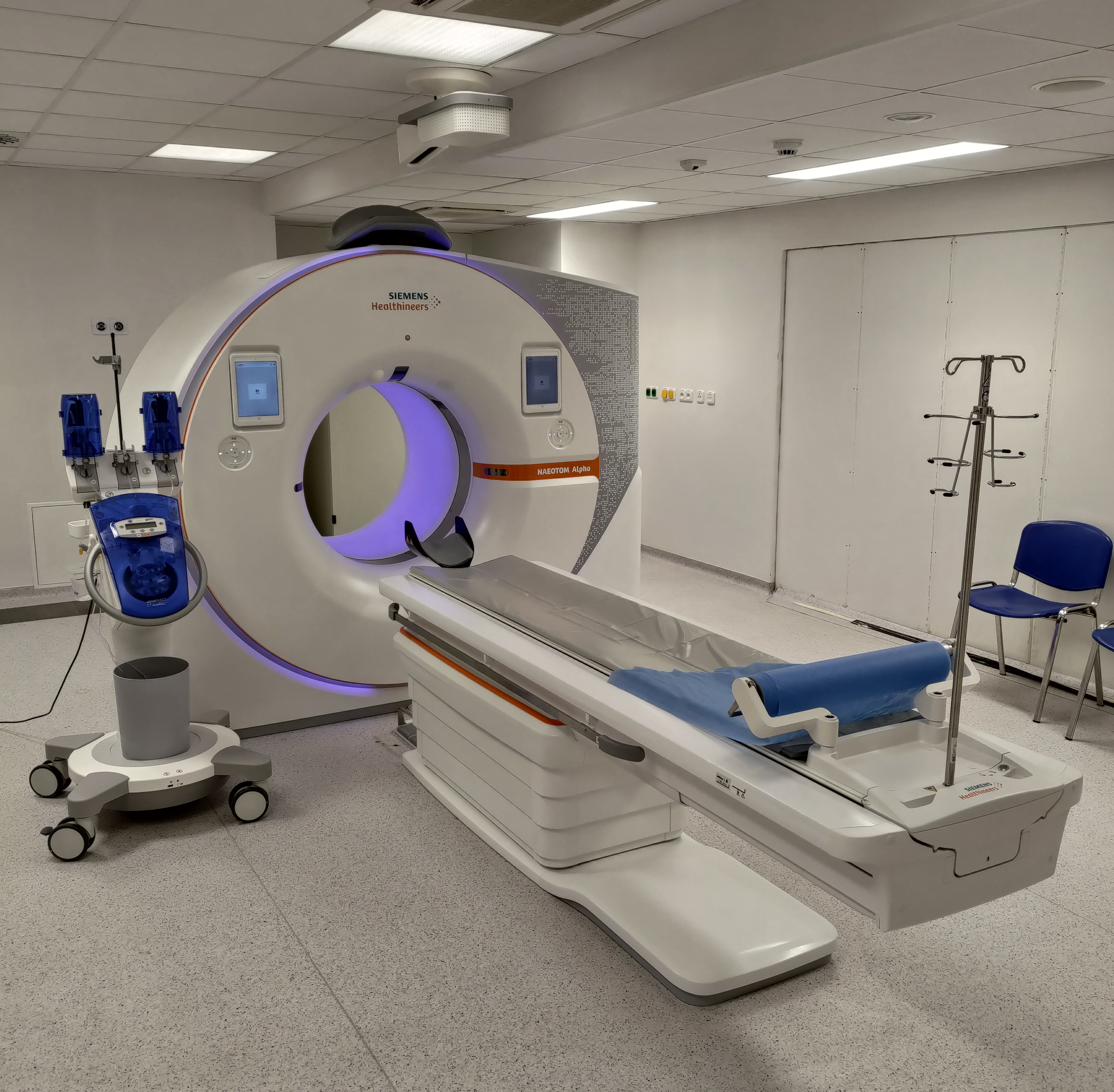}
\caption{Modern CT scanner located at the Lochotín University Hospital in Pilsen, Czech Republic~\cite{109}.}
\label{fig:ctscanner}
\end{figure}

Computed tomography (CT) scan is a medical imaging technique that uses a combination of X-ray equipment and computer technology to produce cross-sectional images, both horizontally and vertically of the body. Compared to x-ray images, these cross-sectional images provide better detail~\cite{7}. CT scan turns two-dimensional X-ray images into three-dimensional ones to obtain more information. Specialists examine the images after getting them to determine the patient's condition. Radiologists can more quickly identify conditions like cancer, cardiovascular disease, infectious disease, trauma, and musculoskeletal diseases by using CT scans. In 1979, South African-American physicist Allan MacLeod Cormack and British electrical engineer Godfrey Hounsfield were awarded Nobel Prize in Physiology or Medicine for the development of computer tomography (CT)~\cite{8}. Figure~\ref{fig:ctscanner} shows a modern CT scanner.

\begin{figure}
\centering
\includegraphics[height=5cm]{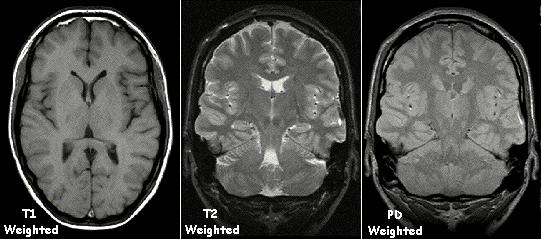}
\caption{Examples of T1-weighted, T2-weighted, and PD-weighted MRI scans~\cite{112}.}
\label{fig:mri_scanned_image}
\end{figure}

In the late 1970s, physicists Peter Mansfield and Paul Lauterbur introduced an MRI-related technique, like the echo-planar imaging (EPI) technique~\cite{mansfield1975diffraction}. Then in 1971 at Stony Brook University, Paul Lauterbur applied magnetic field gradients in all three dimensions and a back-projection technique to create NMR images and published the first images of two tubes of water in the journal Nature~\cite{lauterbur1973image}. The non-invasive medical imaging procedure known as magnetic resonance imaging (MRI) creates detailed images of your body's organs and tissues by combining radio waves generated by a computer and a magnetic field. An MRI machine may detect water molecules inside a body when it is placed into the machine. Water is distributed all over the body. MRI machines can differentiate and construct an image by passing radio waves throughout the body. MRI provides better quality images compared to CT scans for the diagnosis of a disease. Figure~\ref{fig:mri_scanned_image} represents an MRI-scanned image.

Ultrasound is a type of medical imaging that creates images of your body's organs, tissues, and other structures using high-frequency sound waves. It is also known as sonography. It generates a sound wave above 20KHz. Ultrasound can be used in different ways such as finding the unborn baby’s position and condition, abnormality in blood rate, the problem with the structure of the heart, any blockage in the gallbladder, etc. Because the sound waves that ultrasound produces are not very hazardous, it is a safe diagnostic procedure. There are some limitations to using ultrasound. Since sound waves can’t pass through air and bone, ultrasound is ineffective at imaging body parts with gas in them or those covered by bone. To view and examine such areas CT scan, MRI, or X-ray is being used. Figure~\ref{fig:ultrasound_unborn} represents an ultrasound image of an unborn baby.

\begin{figure}
\centering
\includegraphics[height=6.2cm]{ultrasound.jpg}
\caption{Ultrasound image (sonogram) of a fetus in the womb, viewed at 12 weeks of pregnancy (bidimensional scan)~\cite{13}.}
\label{fig:ultrasound_unborn}
\end{figure}

\section{Related Works}
The use of medical imaging to identify and diagnose diseases is expanding quickly. Scientists have already explored a variety of methods to accurately identify an illness. Radiography using X-rays is essential in the diagnosis of diseases like cancer, pneumonia, and COVID-19 \cite{nasim2021prominence}. Chest X-ray images have even less variance and little information. They also have low contrast. That’s why image enhancement plays a vital role to extract more information from a low-quality image. Sharma et al.~\cite{sharma2022comparative} proposed some image enhancement techniques such as Contrast-limited adaptive histogram equalization (CLAHE), decorrelation stretch, morphological operations, median filtering, and noise reduction techniques like a median filter, DCT, and DWT. A total of 6334 chest X-ray images were evaluated by the authors using these methods. Then enhanced images are evaluated by different image quality assessment parameters such as MSE, PSNR, and AMBE. The optimum outcome is obtained by combining the CLAHE and DWT techniques. Meenakshi et al.~\cite{meenakshi2022pneumonia} proposed another computer-aided diagnostic method of Gaussian filter, CLAHE, and Sobel edge detection. The authors applied this strategy to X-ray images of pneumonia. Among women, breast cancer is a prevalent form of cancer. Lu et al.~\cite{lu2017novel} developed a computer-aided diagnosis system for breast MRI. Some feature extraction techniques such as morphological, Gabor feature, etc are applied to the dataset. The classification of breast cancer is then done using an ensemble learning technique that uses weighted averages and a majority vote.

The medical industry is always evolving in terms of technology and procedures. In the past, researchers have attempted to collect more information by enhancing low-quality images with high-quality ones using conventional techniques. They used several image enhancement methods in medical imaging, such as CLAHE, histogram equalization, morphological procedures, etc. With this traditional method, information must be manually extracted. Several statistical features such as bias, variance, mean, perimeter, etc can be applied to medical imaging data to identify a disease. However, researchers are now experimenting with new technologies that eliminate the requirement for manual feature extraction. Deep learning is a technology where features are extracted automatically.

Albahli et al.~\cite{albahli2021coronavirus} published work on COVID-19 using chest X-ray images. A transfer learning model is developed with a 5-fold cross-validation technique. Three models were presented by the author: DenseNet, InceptionV3, and Inception-ResNetV4. The DenseNet model has the highest classification accuracy (92\%), followed by InceptionV3 (83.47\%) and Inception-ResNetV4 (85.57\%). Chakraborty et al.~\cite{chakraborty2020pneumonia} proposed 7 layers of the Convolutional Neural Network (CNN) model. The author used chest X-ray images of children aged 1 to 5 to identify Pneumonia and also utilized optical coherence tomography to identify eye problems (OCT). Sourab et al.~\cite{sourab2022comparison} suggested a novel hybrid approach. They proposed a CNN architecture with 22 layers to extract features from chest X-ray images for Pneumonia. Then some machine learning algorithms such as Random Forest (RF), k-nearest neighbors (KNN), and support vector machine (SVM) are utilized to classify Pneumonia. CNN-RF hybrid model provides the best-performing result of 99.82\% accuracy and 98.7\% AUC. The ensemble is another technique where more than two models are utilized to increase the model’s performance. 

Ravi et al.~\cite{ravi2023multichannel} developed a stacking ensemble technique to detect lung disease from chest X-ray images. First, they used EfficientNetB0, EfficientNetB1, and EfficientNetB2 pre-trained models to extract features. Then the extracted features are combined and a non-linear fully connected layer is included. After that, a stacking ensemble technique is applied to classify lung disease. In a stacking ensemble, random forest and SVM are used in the first stage and logistic regression in the second stage. The accuracy of the proposed method for detecting pediatric pneumonia lung disease, TB lung disease, and COVID-19 lung disease was 98\%, 99\%, and 98\% respectively.
Sejuti et al.~\cite{sejuti2023hybrid} proposed a hybrid CNN-KNN approach for the identification of COVID-19 from computer tomography (CT) scan images. The proposed method is used for a dataset that includes 4085 more CT scan images. After performing 5-fold cross-validation, the proposed method's average accuracy, precision, recall, and F\textsubscript{1} score are 98.26\%, 99.42\%, 97.2\%, and 98.19\%, respectively. Gao et al.~\cite{gao2017classification} proposed a CNN-based model for the diagnosis of Alzheimer’s disease (AD) CT images. They developed a 2D and 3D CNN architecture for detecting the disease. The image dataset is divided into three categories: AD, lesions, and normal. An accuracy of 87.6\% is offered by the proposed methodology. A deep neural network was created by Jalali et al.~\cite{jalali2021resbcdu} to segment lung CT images automatically. They utilized ResNet-34 architecture and BConvLSTM (Bidirectional Convolutional Long Short-term Memory) as an advanced integrator module. The suggested approach yields a dice coefficient index of 97.31\%. Schmauch et al.~\cite{schmauch2019diagnosis} proposed a deep learning-based transfer learning model for the diagnosis of focal-level lesions. They employed ResNet50 to extract features after removing the final two layers. To classify the disease, logistic regression classifier is used which ranges from 0 to 1. A weighted mean ROC-AUC score of 0.891 is achieved by the model. 

Pourasad et al.~\cite{pourasad2021presentation} developed a novel architecture for diagnosis and identifying breast cancer locations using ultrasound images. The fractal approach is used to extract features from images. Images are classified using KNN, SVM, decision tree, and Naive Bayes. Then a convolutional neural network is developed to classify breast cancer images. In the validation, the sensitivity was found 88.5\% and the accuracy for the training set was found 99.8\%. A morphological operation is performed to locate the tumor’s location and volume from the image data. 

Technology advances and federated learning make it possible for healthcare organizations to train machine learning models with private data without compromising 
patient confidentiality through federated learning~\cite{talukder2022federated,puppala2022towards}.
Hossain et al.~\cite{hossain2023collaborative} propose a collaborative federated learning system that enables deep-learning image analysis and classifying diabetic retinopathy without
transferring patient data between healthcare organizations. Along with image data, healthcare patients' statistical data can also be used to train
machine learning models in order to predict disease exposure~\cite{talukder2022prediction,puppala2023machine}.

\begin{table}[!h]
\caption{Related work in a nutshell}
\label{T:Ley de Kirchhoff}
\begin{center}
\begin{tabular}{| p{1.8cm} | p{5.5cm} | p{2.6cm}| p{3.4cm} |} 
    \hline

     \textbf{Author} & \textbf{Methodology} & \textbf{Disease} & \textbf{Performance} \\
    \hline

 Sharma et al.~\cite{sharma2022comparative} & Contrast-limited adaptive histogram equalization (CLAHE), Decorrelation stretch, Morphological operation, Median filter, DCT, and DWT & Chest X-ray images & - \\ 
      \hline
Meenakshi et al.~\cite{meenakshi2022pneumonia} & Gaussian filter, CLAHE, and Sobel edge detection & X-ray images of Pneumonia & -   \\ 
      \hline
Lu et al.~\cite{lu2017novel} & Morphological operation, Gabor filter and Ensemble learning & Breast MRI & ROC-AUC score – 0.9617  \\ 
      \hline
 
Albahli et al.~\cite{albahli2021coronavirus} & 5-fold cross-validation technique, Transfer Learning technique: DenseNet, InceptionV3, and Inception-ResNetV4 & COVID-19 chest X-ray images  &Accuracy - 92\% \\ 
      \hline   
Chakraborty et al.~\cite{sourab2022comparison} & Convolutional Neural Network (CNN) & Chest X-ray of Pneumonia & accuracy – 63\%  \\ 
      \hline   
Sourab et al.~\cite{ravi2023multichannel} & CNN with machine learning algorithms: Random Forest (RF), K-Nearest Neighbors (KNN), and Support Vector Machine (SVM) & Chest X-ray images of Pneumonia  & Accuracy - 99.82%
AUC - 98.7\%  \\ 
      \hline   
Ravi et al.~\cite{sejuti2023hybrid} & Stacking ensemble technique: EfficientNetB0, EfficientNetB1, EfficientNetB2,
RF and SVM and Logistic regression.
 & Pneumonia lung disease, TB lung disease, and COVID-19 lung disease  & Pneumonia lung disease (accuracy) – 98\%
TB lung disease
COVID-19 (accuracy) – 99\%
Lung disease (accuracy) – 98\%  \\ 
      \hline   
Sejuti et al.~\cite{gao2017classification} & Hybrid CNN-KNN & COVID-19 from computer tomography (CT) scan images  & Accuracy – 98.26\%
Precision – 99.42\%
Recall – 97.2\%
F\textsubscript{1} score – 98.19\%
 \\ 
      \hline  
Gao et al.~\cite{jalali2021resbcdu} & CNN architecture & Alzheimer’s disease (AD) CT images. & Accuracy – 87.6\%  \\ 
      \hline

      \hline  
Jalali et al.~\cite{schmauch2019diagnosis} & ResNet-34 architecture and BConvLSTM (Bidirectional Convolutional Long Short-term Memory) & Lung CT images & Dice coefficient index – 97.31\%.  \\ 
     
            \hline  
Schmauch et al.~\cite{pourasad2021presentation} & ResNet50 and Logistic regression & Focal-level lesions & AROC-AUC score – 0.891 \\ 
     
            \hline  
Pourasad et al.~\cite{guan2014anisotropic} & CNN with KNN, SVM, Decision Tree, and Naive Bayes and morphological operation & Breast cancer locations using ultrasound images & Sensitivity – 88.5\%
Accuracy (training set) – 99.8\%.
  \\ 
      \hline

\end{tabular}
\end{center}
\end{table}

\section{Methodology}
Computer-aided diagnostic systems are widely used by researchers to identify any disease. In the medical field, there are many ways to detect the disease, of which Machine Learning and Deep Learning are widely used nowadays.

\subsection{Image Processing}
Image processing is the initial stage of disease detection in medical imaging. The extraction of more information from an image is the main objective of image processing. The more information you obtain, the easier it will be to detect any flaw in medical imaging. There are several image processing techniques available such that denoising, image enhancement, segmentation, morphological operation, etc. Medical imaging like CT and MRI scans have noisy elements that reduce the image's overall effectiveness. Image denoising techniques reduce noise and improve the amount of information in the image that is produced. Common techniques to remove noise from medical images include Gaussian averaging, mean, median~\cite{guan2014anisotropic}, Lee~\cite{lee1980digital}, and diffusion filters~\cite{yu2002speckle}. These filters and their variations are created to eliminate particular types of noise in various medical imaging modalities~\cite{mittal2010enhancement,li2012sar,abd2002real}. These filters typically perform a low pass filtering by eliminating the distinct peaks and replacing the suspected values with a local average or other locally relevant measurements. 

\begin{figure}
\centering
\includegraphics[height=5cm]{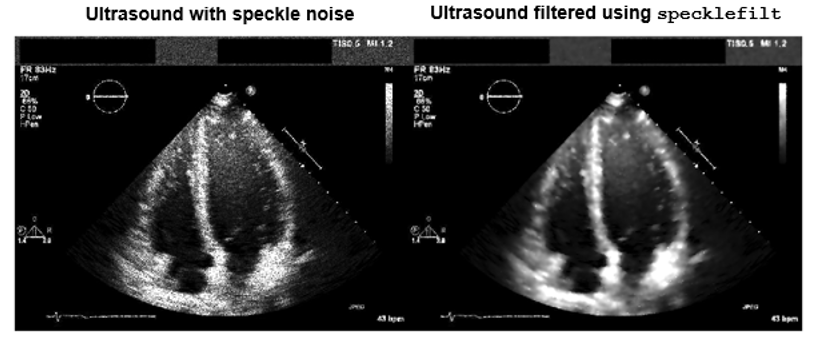}
\caption{Image denoising~\cite{abd2002real}.}
\label{fig:denoising}
\end{figure}

\begin{figure}
\centering
\includegraphics[height=5cm]{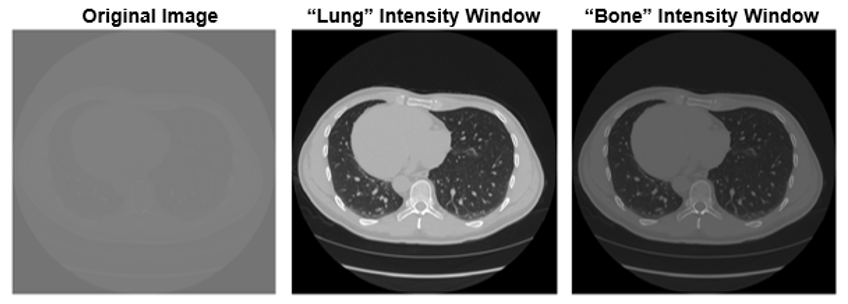}
\caption{Image Enhancementl ~\cite{abd2002real}.}
\label{fig:enhancement}
\end{figure}

Image enhancement includes making adjustments to digital images to make them more acceptable for display or additional image analysis. Image enhancement enhances the visual quality of the image, supports the clinician's selection, and ultimately protects the lives of the patients. There are some useful examples and methods of image enhancement: 1. Morphological operations, 2. Linear contrast adjustment, 3. Histogram equalization, 4. Contrast-limited adaptive histogram equalization (CLAHE) 5. Wiener filter, 6. Median filtering and 7. Decorrelation stretch. Medical image segmentation is the procedure used to separate regions of interest (ROIs) from image data, such as that from CT or MRI scans. Identifying the anatomical regions required for particular research is the main goal of segmenting this data. One of the key benefits of medical image segmentation is that separating just essential regions allows for more precise anatomical data interpretation. Segmentation also has the advantage of removing any irrelevant information from a scan, such as air and enabling the separation of various tissues, such as bone and soft tissues. Edge-based, region-based, thresholding approaches and other types of medical image segmentation methods are available. Figures~\ref{fig:denoising},~\ref{fig:enhancement},~\ref{fig:image_segmentation} represent some image processing techniques.

\begin{figure}
\centering
\includegraphics[height=5cm]{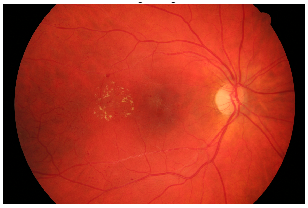} 
\includegraphics[height= 5 cm]{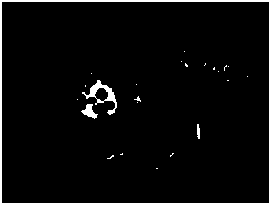}
\caption{Original Image (left) and Image Segmentation (right)}
\label{fig:enhancementseg}
\end{figure}


\subsection{Machine Learning}

Machine Learning is a subset of Artificial Intelligence (AI) where a machine can act like a human and speed up the learning process to provide a better result. Machine learning supports manual feature extraction techniques. It first extracts the feature from the images before sending them to the supervised learning classifier to classify an object. It is necessary to choose several feature types when all image processing has been completed. Features are numerical numbers that help with further classification. There are two different kinds of statistical features: wavelet features and GLCM features. These GLCM features are Autocorrelation, Correlation, Dissimilarity, Cluster Prominence, Contrast, Cluster Shade, Energy, Entropy1, Homogeneity1, Homogeneity2, Maximum probability, Sum of squares, Sum entropy, Sum average, Sum variance, Difference variance, Difference entropy, etc. The texture features are Area, Mean, Standard Deviation, Entropy2, RMS, Kurtosis, Skewness, Variance, IDM, and Smoothness. In this method, features are manually extracted. There are a certain amount of features available in the manual feature selection process. These features are applied to the machine learning algorithm to classify diseases.

\subsection{Deep Learning}
Deep learning is a branch of machine learning that simulates the behavior of a neuron. There are several types of deep learning methods available such that custom CNN, transfer learning models, custom CNN with machine learning algorithm (CNN-ML), etc. Deep learning can extract features automatically from input images \cite{tonmoy2019brain}. This is the component of the method that matters the most. It is capable of automatically picking up complex patterns and characteristics from objects. Deep learning outperforms machine learning in terms of performance because of this automatic feature extraction technique. When deciding to use deep learning rather than machine learning, a powerful GPU and a ton of data are required. Deep learning would not be the best option if there is any lag. Some of the deep learning approaches are given below.
\subsubsection{Custom CNN}
A deep learning model of custom CNN is built by adding layers from scratch. There are different types of layers are available to develop a custom CNN model. These are the convolutional layer, activation layer, batch normalization layer, etc. Every layer has a specific parameter, and these parameters must be adjusted to achieve optimum results. The main objective of a convolution layer is to extract features from an image. It's necessary to specify the kernel size in the convolutional layer. Kernel size describes the amount of extracted features. It's also necessary to provide the filter's size. Stride is another parameter that is applied to compress the size of the images. The number of pixels that are shifted across the input matrix is known as the stride. Padding is the technique of enhancing the input image with additional pixel values. Zero padding is the addition of zero pixels to the input image's border when the padding value is set to zero. Padding makes the input matrix larger to improve the accuracy of the analysis. The activation function is what the activation layer consists of. If neurons are active or not, is determined by a mathematical process known as the activation function. There are several activation functions available such as Sigmoid, Tanh, ReLu, Softmax activation function, etc. When the data is provided o the model, then weights and biases make the data linear. For a neuron, learning linear data is quite straightforward. However, real-world data is more complicated. The activation function transforms linear data into non-linear data. It assists neurons in learning more complex patterns. The most common activation function is the step function which is a threshold. Neurons become active at a particular value, and they become inactive below a given threshold. Step-function is mainly used for binary classification. However, it is impossible to determine multi-classification. The zero gradient descent value is a limitation of the step function. As a result, weights cannot be updated. No improvement will be added to the model's performance. Softmax is another type of activation function that is used in the classification layer of a neural network. It is used in multinomial logistic regression and normalizes the output of a network to a probability distribution over the predicted output class. The formula for the softmax activation function is

\begin{equation}
    f\left ( \sum_{i=1}^{n} x_{i}w_{i}+b\right )
\end{equation}

\begin{figure}
\centering
\includegraphics[height=6.2cm]{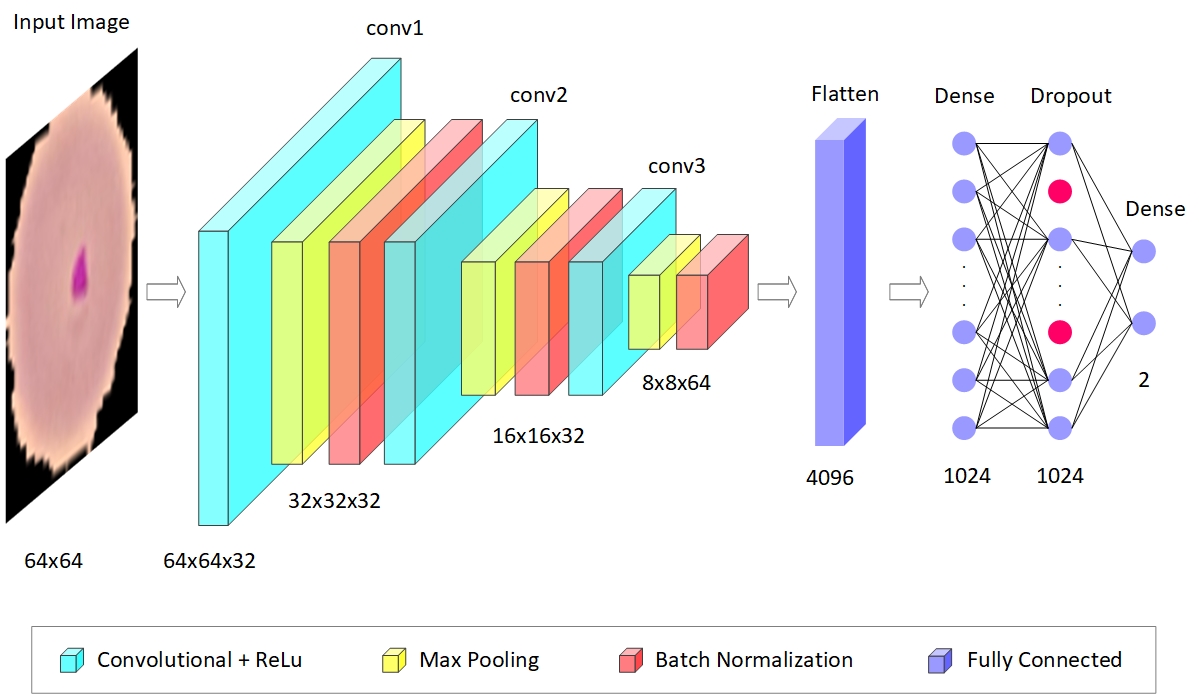}
\caption{Custom CNN architecture~\cite{32}.}
\label{fig:customCNN_architecture}
\end{figure}

Where,
	b = bias
	x = input to a neuron
	w = weights
	n = number of inputs from the incoming layer
	i = counter from 1 to n
Another element that makes up CNN is the pooling layer. It reduces the number of parameters and computation of the CNN model. Some examples are max pooling and average pooling. Maximum values are extracted from a rectified feature map using max pooling. The average operation is carried out to take average values using average pooling. Up to this point, the convolutional layer, which extracts features from images, has been described. It's time to categorize the image at this point. The classification function is carried out via a fully connected layer. The feed-forward neural network and the fully connected layer both operate similarly. The convolutional layer contains a three-dimensional array of data. The data is delivered as a one-dimensional array to a fully connected layer. The flattened layer reduces data from three dimensions to one dimension. This is the first fully connected layer.
How many layers to utilize will depend on the application. It is possible to choose the required number of neurons for that layer. The classification layer is the topmost and last layer. The required output determines how many neurons are needed in the bottom layer. For example, if we wish to classify both dogs and cats, the last layer will have 2 neurons. Some hyperparameters are employed to achieve the best results. Some of the hyperparameters include input size, batch size, epoch, learning rate, optimizer, etc. Figure~\ref{fig:customCNN_architecture} shows a custom CNN architecture \cite{hossainbrain}.

\subsubsection{Transfer Learning}
\begin{figure}
\centering
\includegraphics[height=6.2cm]{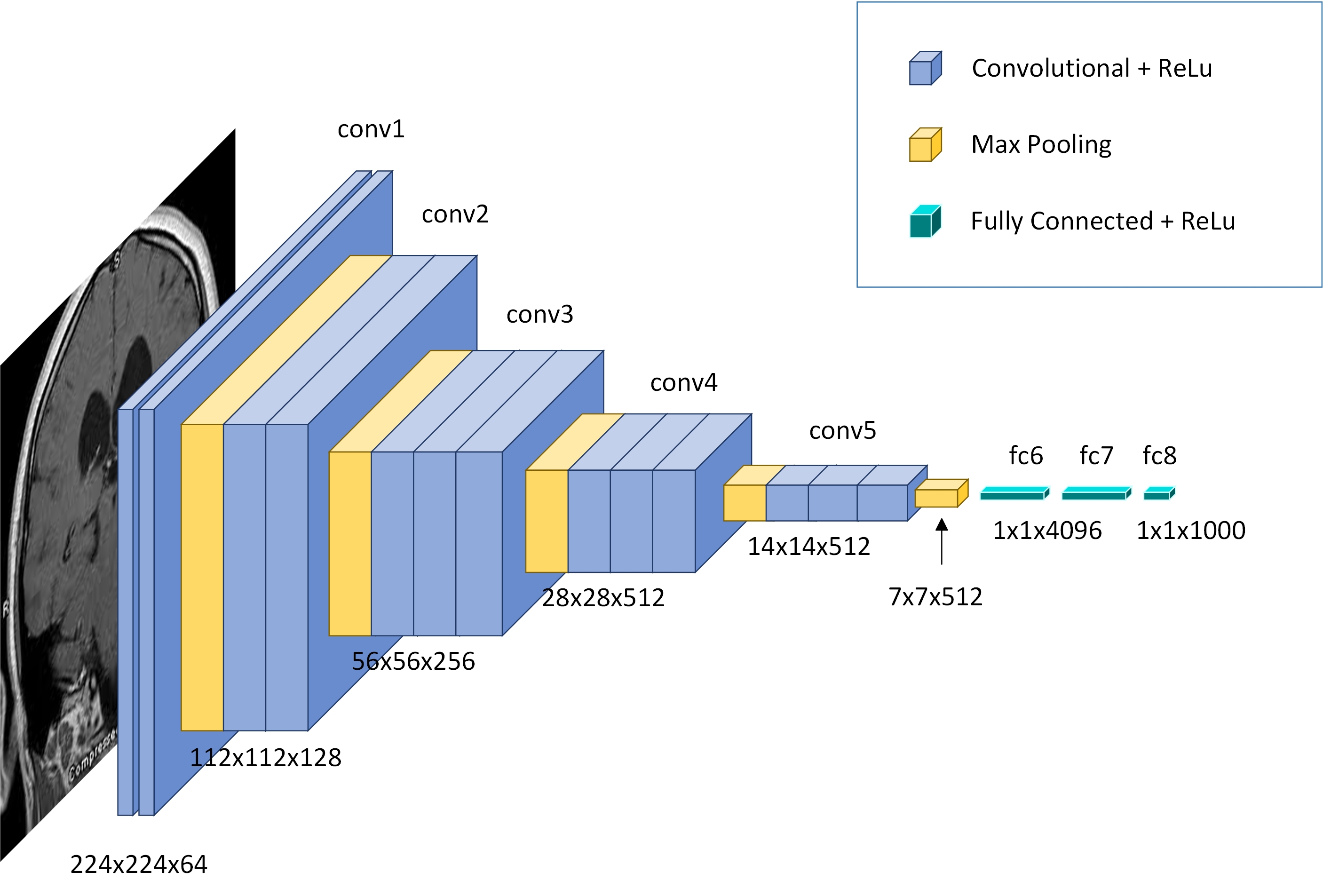}
\caption{VGG16 architecture~\cite{simonyan2014very}.}
\label{fig:vgg16_architecture}
\end{figure}

Transfer learning is a pre-trained network where a model is trained for one task and can be reused in another task. Researchers already developed different transfer learning models for multiclass classification. Several pre-trained networks are available for example AlexNet~\cite{krizhevsky2017imagenet}, VGG16~\cite{simonyan2014very}, ResNet50~\cite{he2016identity}, etc. Figure~\ref{fig:vgg16_architecture} represents the VGG16 architecture.

\begin{figure}
\centering
\includegraphics[height=6.2cm]{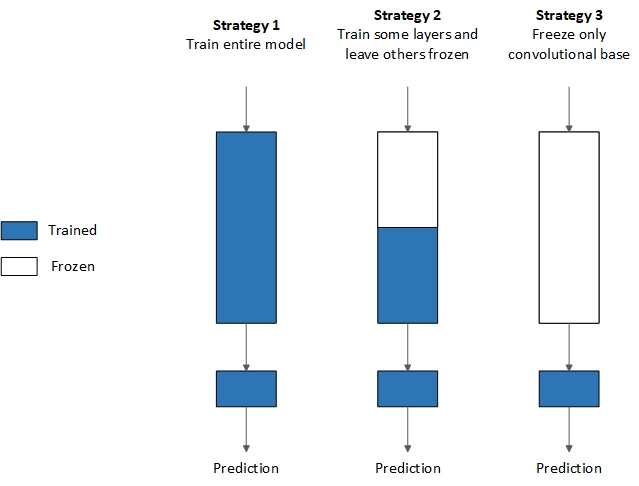}
\caption{Transfer learning strategies.}
\label{fig:transfer_learning_strategies}
\end{figure}

Transfer learning models can be reused by applying different strategies. 
Train the entire model: Transfer learning can be retrained with new weights for a specific application because it is a pre-trained network with pre-existing weights. The architecture would remain unchanged, but a new dataset would be used to update the weights. The final layer must be modified based on how many classes are used in that application. For example, AlexNet~\cite{krizhevsky2017imagenet} was trained on 1000 categories. The model can be trained using five categories. It depends on the application of how many classifications there are.
Freeze some layers and train others: This is an additional method of freezing some layers of the transfer learning model. The weights of the frozen layer will remain constant. The other layers receive training using the existing weights. The quantity of training and freezing layers is adjustable based on the application.
Freeze all layers and change only the final layer: A transfer learning model with existing weights will be employed in this technique. The layers won't be updated with new weights as they are frozen. According to the different classifications, just the top layer will be altered. Strategies of the transfer learning model are shown in Figure~\ref{fig:transfer_learning_strategies}

\subsubsection{CNN-ML}

\begin{figure}
\centering
\includegraphics[height=5.8cm]{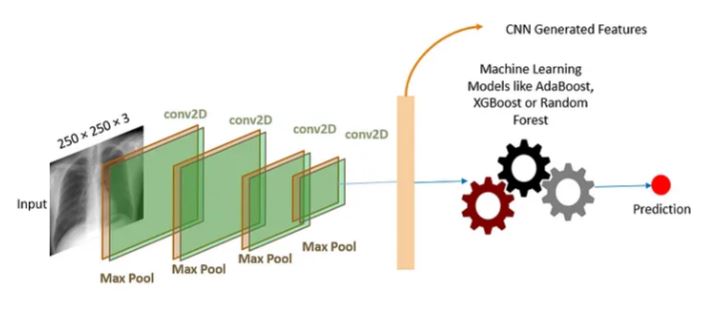}
\caption{Simple architecture for a deep hybrid network model~\cite{deephybridnet}.}
\label{fig:deephybridnet}
\end{figure}

\begin{figure}
\centering
\includegraphics[height=5cm]{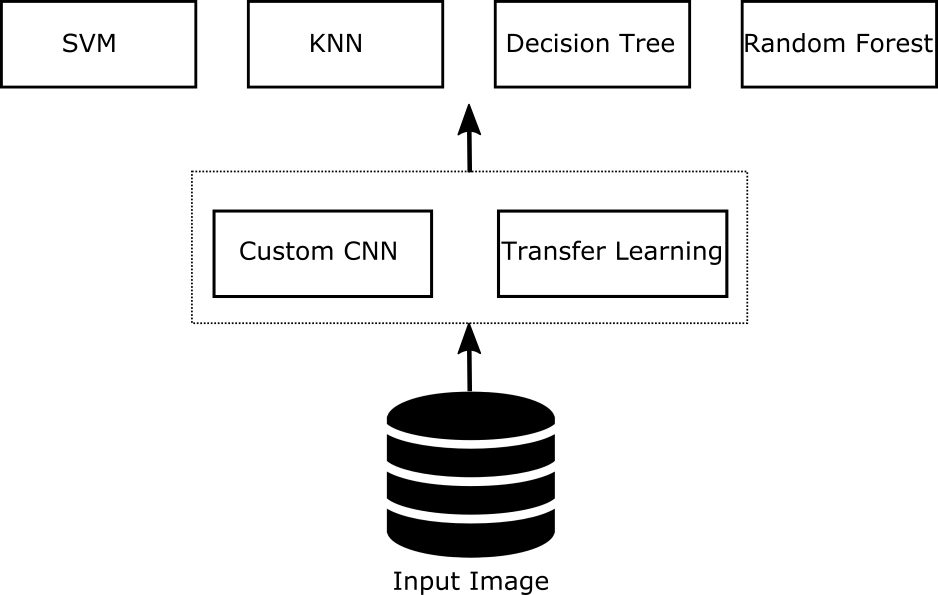}
\caption{CNN-ML architecture.}
\label{fig:CNNML_architecture}
\end{figure}

Now, machine learning algorithms will be utilized as a classifier and CNN will be used as a feature extractor. Any transfer learning model or a custom CNN can be used as the CNN model. SVM, KNN, Decision Tree, and Random Forest are a few machine learning algorithms that can be used as classifiers. An easy Deep Hybrid Network model design is shown in Figure~\ref{fig:deephybridnet}. Here, the DNN layer is composed of only four layers, and the ML classification layer follows. Figure~\ref{fig:CNNML_architecture} represents a CNN-ML architecture. ML classification algorithms like AdaBoost, XGBoost, and Random Forest are used to describe this layer. It is advised to use a more sophisticated DNN layer for more difficult issue-solving and better model performance.

Ketu et al.~\cite{deephybridnet} introduce a CNN-LSTM hybrid deep learning prediction model, which can correctly forecast the COVID-19 epidemic across India. The proposed model uses convolutional layers, to extract meaningful information and learn from a given time-series dataset. 
By fusing deep learning with machine learning, a fusion network known as deep hybrid learning may be created. Deep hybrid learning, in which the model produces or extracts features from unstructured data using deep learning techniques and then utilizes traditional machine learning techniques to create highly accurate classification models utilizing the unstructured data. So, combining DL and ML with Deep Hybrid Learning (DHL) may improve upon each approach's shortcomings while also delivering more accurate and computationally efficient results.

\section{Performance Analysis}
There are various evaluation metrics available both for regression and classification tasks. Some of the performance metrics for medical image classification are Accuracy, Precision, Recall, F1-Score, Confusion Matrix, ROC (Receiver Operating Characteristic Curve), AUC (Area Under the ROC Curve) curve, etc. Accuracy is the number of correct predictions to the number of total predictions. The Accuracy measure should be utilized when the target variable classes in the data are fairly equal. It's suggested against using the accuracy measure when the target variable primarily belongs to one class. Precision is another performance metric which is the ratio of correctly positive observations to the total predicted positive observations. Accuracy has a constraint, which is overcome by the precision metric. It measures the proportion of actual positives that were incorrectly detected and is comparable to the Precision metric. Precision provides information on how well a classifier performs with false positives, whereas recall measures how well a classifier performs with false negatives. The weighted average of recall and precision is the F\textsubscript{1} score. When the class of the dataset is highly imbalanced, then the F\textsubscript{1} score is utilized. 

True positive (TP), false positive (FP), true negative (TN), and false negative (FN) are the four key parameters. True Positive (TP) is a metric for evaluating how many samples your model properly predicted as positive. The amount of negative class samples that your model correctly predicted is known as True Negative (TN). The term False Positive (FP) indicates how many negative class samples your model incorrectly predicted as positive. The number of samples from the positive class that your model incorrectly predicted as negative is represented by False Negative (FN). In equations 2-5, different performance metrics are shown along with a few details~\cite{sharma2022comparative,al2022brain}.

\textbf{Precision} is calculated using the following equation number 2.

\begin{equation}
    Precision = \frac{TP}{TP+FP}
\end{equation}

\begin{figure*}[ht!]
\centering
\includegraphics[width=\textwidth]{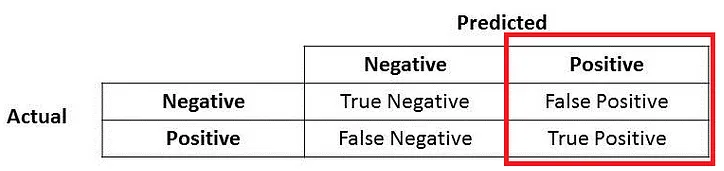}
\caption{Precision.}
\label{fig:confusion_matrix_precision}
\end{figure*}

From Figure~\ref{fig:confusion_matrix_precision} we see the denominator in equation 2 is the total predicted positive. So, we can rewrite the equation as follows:

\begin{equation}
    Precision = \frac{TP}{\text{Total Predicted Positive}}
\end{equation}

We see that Precision tells us how accurate the model is. Out of those predicted positives, how many of them are positive?

\textbf{Recall} can be calculated using the following equation (equation 4).
\begin{equation}
    Recall = \frac{TP}{TP+FN}
\end{equation}

\begin{figure*}[ht!]
\centering
\includegraphics[width=\textwidth]{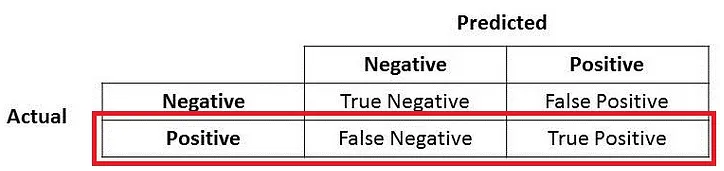}
\caption{Recall.}
\label{fig:confusion_matrix_recall}
\end{figure*}

From Figure~\ref{fig:confusion_matrix_recall} we see the denominator in equation 4 is the total actual positive. So, we can rewrite the equation as follows:

\begin{equation}
    Recall = \frac{TP}{\text{Total Actual Positive}}
\end{equation}

Here we see that Recall calculates the ratio between the number of predicted positives and the number of actual positives.

\begin{table}[]
\centering
\begin{tabular}{cccll}
\cline{1-3}
\multicolumn{1}{|c|}{\textbf{n=165}}                                                  & \multicolumn{1}{c|}{\textbf{\begin{tabular}[c]{@{}c@{}}Predicted: \\ No\end{tabular}}} & \multicolumn{1}{c|}{\textbf{\begin{tabular}[c]{@{}c@{}}Predicted:\\ Yes\end{tabular}}} &  &  \\ \cline{1-3}
\multicolumn{1}{|c|}{\textbf{\begin{tabular}[c]{@{}c@{}}Actual:\\ No\end{tabular}}}   & \multicolumn{1}{c|}{50}                                                                & \multicolumn{1}{c|}{10}                                                                &  &  \\ \cline{1-3}
\multicolumn{1}{|c|}{\textbf{\begin{tabular}[c]{@{}c@{}}Actual: \\ Yes\end{tabular}}} & \multicolumn{1}{c|}{5}                                                                 & \multicolumn{1}{c|}{100}                                                               &  &  \\ \cline{1-3}
\multicolumn{1}{l}{}                                                                  & \multicolumn{1}{l}{}                                                                   & \multicolumn{1}{l}{}                                                                   &  & 
\end{tabular}
\caption{Confusion matrix example}
\label{tab:confusion_matrix_example}
\end{table}

\textbf{F\textsubscript{1} score} is calculated using the following equation (equation 6).
\begin{equation}
    F\textsubscript{1} score= 2\times \frac{Precision\times Recall}{Precision + Recall}
\end{equation}
F\textsubscript{1} score is a function of Precision and Recall. It is calculated from the harmonic mean of the precision and recall. F\textsubscript{1} score ranges from 0 to 1.
F\textsubscript{1} score is also known as balanced F-score or F-measure.

\textbf{Accuracy} is calculated using the following equation (equation 7).
\begin{equation}
    Accuracy = \frac{TP+TN}{TP+FP+TN+FN}
\end{equation}

The ground-truth labels and model predictions are displayed in a table using the Confusion Matrix. In the confusion matrix, the cases in a predicted class are represented in each row, whereas the occurrences in an actual class are shown in each column. A confusion matrix is shown in Table~\ref{tab:confusion_matrix_example}.

\begin{figure}
\centering
\centering
\includegraphics[height=4cm]{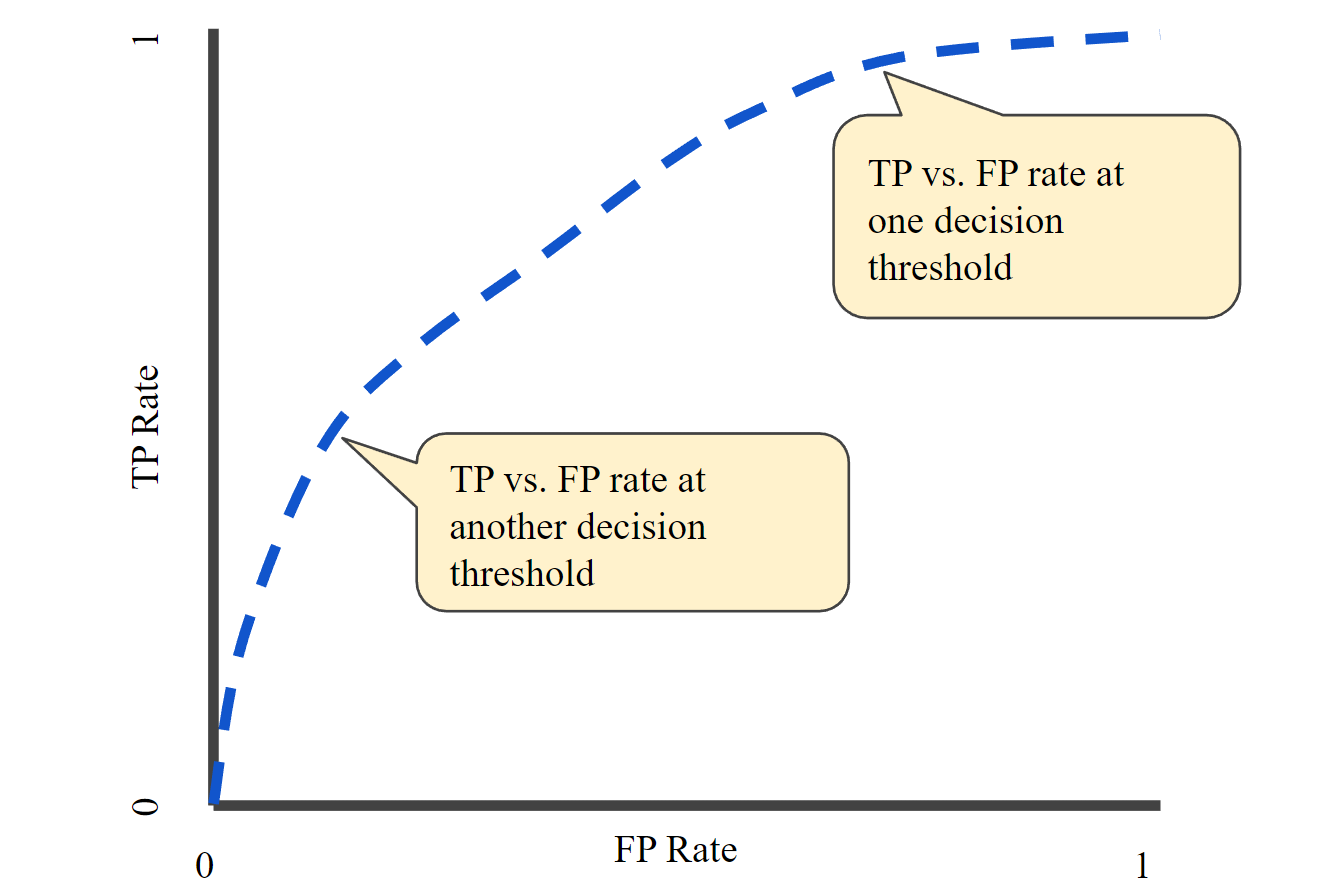}
\caption{ROC curve} \label{fig:roc_curve}
\hfill
\centering
\includegraphics[height=4cm]{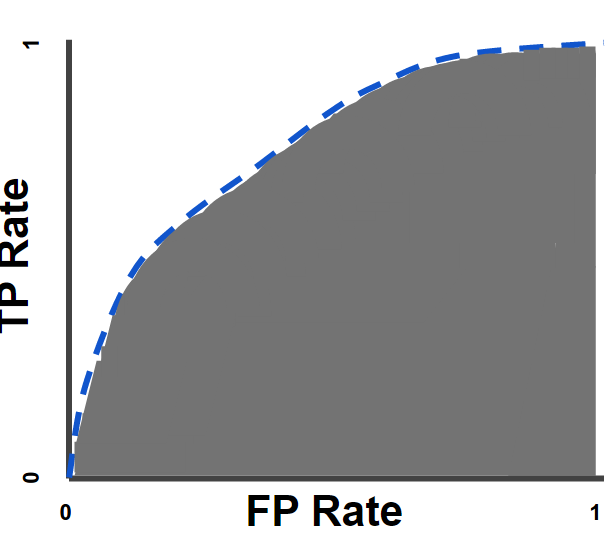}
\caption{AUC curve}
\label{fig:auc_curve}
\caption{ROC and AUC Curve~\cite{googleROCAUC}.}
\label{fig:roc_auc_curve}
\end{figure}

\textbf{ROC curve.} It can be used when it's necessary to visualize the performance of the classification model. It is a widely-used statistic that is important for assessing how well the categorization model is working. The performance of a classification model at various threshold levels is represented graphically by the ROC curve. True positive rate and false positive rate are the two parameters between which the curve is displayed. 


\textbf{AUC curve.} The area Under the ROC curve is referred to as the AUC curve. AUC determines performance across all thresholds and offers an overall measurement. AUC values vary from 0 to 1. AUC 0 means a model with 100\% wrong prediction and AUC 1 means 100\% correct prediction. It assesses the accuracy of the model's predictions without taking the categorization threshold into consideration.

\section{Conclusion}
This study reviews a few research publications and discusses many sorts of methodologies for medical imaging \cite{shahbrain}, such as CT, MRI, and X-ray. Medical imaging has been increasingly popular in recent years for disease diagnosis.  Researchers are looking for novel ways to quickly and precisely diagnose diseases. Radiologists can utilize a variety of computer-aided automatic technologies to find the disease. Feature extraction is made easier by image processing techniques that improve image quality. Machine learning has the ability to extract the features manually. Because of this, the machine learning model performs worse than the deep learning approach. When compared to machine learning, the deep learning model offers the most accurate results and can automatically extract features. Radiologists can make an accurate and quick diagnosis of an illness with the help of a computer-aided system.

\bibliographystyle{unsrt}
\bibliography{source}











\end{document}